\documentclass[12pt,preprint]{aastex}
\usepackage{epsfig,amsfonts,amsmath,amssymb,graphicx,morefloats,appendix,tensor}
\usepackage{color}
\usepackage{natbib}
\begin{document}

\slugcomment{Accepted to ApJ: April 6, 2017}

\title{A Complete ALMA Map of the Fomalhaut Debris Disk
}

\author{Meredith A. MacGregor\altaffilmark{1}, Luca Matr\`a\altaffilmark{2}, Paul Kalas\altaffilmark{3,4}, David J. Wilner\altaffilmark{1}, Margaret Pan\altaffilmark{5}, Grant M. Kennedy\altaffilmark{2}, Mark C. Wyatt\altaffilmark{2}, Gaspard Duchene\altaffilmark{3,6}, A. Meredith Hughes\altaffilmark{7}, George H. Rieke\altaffilmark{8}, Mark Clampin\altaffilmark{9}, Michael P. Fitzgerald\altaffilmark{10}, James R. Graham\altaffilmark{3}, Wayne S. Holland\altaffilmark{11}, Olja Pani\'{c}\altaffilmark{12}, Andrew Shannon\altaffilmark{13,14,2}, Kate Su\altaffilmark{8}}

\altaffiltext{1}{Harvard-Smithsonian Center for Astrophysics, 60 Garden Street, Cambridge, MA 02138, USA}
\altaffiltext{2}{Institute of Astronomy, University of Cambridge, Madingley Road, Cambridge CB3 0HA, UK}
\altaffiltext{3}{Astronomy Department, University of California, Berkeley CA 94720-3411, USA}
\altaffiltext{4}{SETI Institute, Mountain View, CA, 94043, USA}
\altaffiltext{5}{MIT Department of Earth, Atmospheric, and Planetary Sciences, Cambridge, MA 02139, USA}
\altaffiltext{6}{Universit\'e Grenoble Alpes/CNRS, Institut de Plan\'etologie et d'Astrophysique de Grenoble, 38000 Grenoble, France}
\altaffiltext{7}{Department of Astronomy, Van Vleck Observatory, Wesleyan University, Middletown, CT 06459, USA}
\altaffiltext{8}{Steward Observatory, University of Arizona, Tucson, AZ 85721, USA}
\altaffiltext{9}{NASA Goddard Space Flight Center, Greenbelt, MD 20771, USA}
\altaffiltext{10}{Department of Physics and Astronomy, UCLA, Los Angeles, CA 90095}
\altaffiltext{11}{UK Astronomy Technology Centre, Royal Observatory, Blackford Hill, Edinburgh EH9 3HJ, UK; Institute for Astronomy, Royal Observatory, University of Edinburgh, Blackford Hill, Edinburgh EH9 3HJ, UK}
\altaffiltext{12}{School of Physics and Astronomy, University of Leeds, Leeds LS2 9JT, UK}
\altaffiltext{13}{Department of Astronomy \& Astrophysics, The Pennsylvania State University, State College, PA 16801, USA}
\altaffiltext{14}{Center for Exoplanets and Habitable Worlds, The Pennsylvania State University, State College, PA 16802, USA}

\begin{abstract}

We present ALMA mosaic observations at 1.3~mm (223~GHz) of the Fomalhaut system with a sensitivity of 14~$\mu$Jy/beam.  These observations provide the first millimeter map of the continuum dust emission from the complete outer debris disk with uniform sensitivity, enabling the first conclusive detection of apocenter glow.  We adopt a MCMC modeling approach that accounts for the eccentric orbital parameters of a collection of particles within the disk. The outer belt is radially confined with an inner edge of $136.3\pm0.9$~AU and width of $13.5\pm1.8$~AU.  We determine a best-fit eccentricity of $0.12\pm0.01$. Assuming a size distribution power law index of $q=3.46\pm 0.09$, we constrain the dust absorptivity power law index $\beta$ to be $0.9<\beta<1.5$. The geometry of the disk is robustly constrained with inclination $65\fdg6\pm0\fdg3$, position angle $337\fdg9\pm0\fdg3$, and argument of periastron $22\fdg5\pm4\fdg3$.  Our observations do not confirm any of the azimuthal features found in previous imaging studies of the disk with HST, SCUBA, and ALMA.  However, we cannot rule out structures $\leq10$~AU in size or which only affect smaller grains. The central star is clearly detected with a flux density of $0.75\pm0.02$~mJy, significantly lower than predicted by current photospheric models.  We discuss the implications of these observations for the directly imaged Fomalhaut b and the inner dust belt detected at infrared wavelengths.

\end{abstract}


\section{Introduction}
\label{sec:intro}

The proximity of Fomalhaut \cite[$7.66\pm0.04$~pc,][]{van07} has resulted in its debris disk being one of the best-studied.  With an age of $\sim440$ Myr \citep{mam12}, Fomalhaut is at a stage when significant dynamical activity can still occur, as indicated by the period of Late Heavy Bombardment in our own Solar System, an epoch that has important implications for the final architecture of the planetary system.  The outer debris disk is located at $\sim140$~AU, and has been resolved at a range of wavelengths spanning from optical to radio \citep{hol98,kal05,kal08,kal13,acke12,ric12,bol12,whi16}.  In addition to the cold ($\sim50$~K) outer belt, the system has a warm ($\sim150$~K), unresolved inner component detected as excess emission at infrared wavelengths with both \emph{Spitzer} and \emph{Herschel} \citep{sta04,su13}.  \cite{su16} placed limits on the radial location of this inner belt between $\sim8-15$~AU with a non-detection from the Atacama Large Millimeter/submillimeter Array (ALMA).  Direct imaging has also revealed the presence of a very low mass object, Fomalhaut b, near the outer disk and with a highly eccentric orbit \citep{kal08,kal13}.  Given its unique characteristics and architecture, the Fomalhaut system is a Rosetta stone for understanding the interaction between planetary systems and debris disks studying which will enhance our physical understanding of more distant planetary systems.

Dusty debris disks, like the Fomalhaut system, are produced from the continual collisional erosion of larger planetesimals, similar to asteroids or comets. The resulting dust is shaped by the larger bodies or planets in the system through collisions and gravitational perturbations, imprinting observable signatures in the structure of the disk. For example, an interior planet on an eccentric orbit can impose a forced eccentricity on the dust particles in the disk \citep{wya99}.  Such a planet could also sculpt a sharp interior edge \citep{qui06,chi09}.  The outward migration of a planet can radially confine the belt between resonances \citep{hahn05}, similar to Neptune in our own Solar System, or trap dust into mean motion resonance outside its orbit \citep{kuc03,wya03,del05}.  Observations at millimeter wavelengths offer an advantage for probing these planetary-induced structures, since the large grains that emit predominantly at these wavelengths are not significantly perturbed by radiation forces and better trace the location of the larger planetesimals.  Previous resolved images have revealed that the Fomalhaut debris disk is both radially confined and significantly eccentric.  However, there has yet to be a complete map of the disk structure at millimeter wavelengths, necessary to probe for azimuthal disk structure that might stem from planetary interactions.

Here, we present new mosaic observations with the Atacama Large Millimeter/submillimeter Array (ALMA) of the Fomalhaut debris disk, which build the first complete millimeter map of the system at the current epoch.  By mapping the outer belt with uniform sensitivity, we are able to place constraints on the azimuthal structure of the belt and make the first robust observational detection of apocenter glow. In Section~\ref{sec:obs} we present the new ALMA observations.  In Section~\ref{sec:results} we discuss the structure of the continuum emission (\ref{sec:cont}), our modeling approach (\ref{sec:model}), and the results of our modeling (\ref{sec:fits}).  In Section~\ref{sec:disc}, we discuss the significance of the results in the context of apocenter glow (\ref{sec:apo}), the structure and geometry of the disk (\ref{sec:struc}), implications for the directly imaged Fomalhaut b (\ref{sec:fomb}), and constraints on the emission of the central star (\ref{sec:star}). Section~\ref{sec:conc} presents our conclusions.

\section{Observations}
\label{sec:obs}

We observed the Fomalhaut system with ALMA in Band 6 (1.3~mm, 223~GHz).  To map the entire outer dust belt, we constructed a seven pointing mosaic covering the star and the disk circumference.  The phase center for the central pointing was $\alpha=22^\text{h}57^\text{m}39.449$, $\delta=-29\degr37\arcmin22\farcs687$ (J2000), corresponding to the position of the star corrected for its proper motion (328.95, -164.67) mas yr$^{-1}$.  One pointing was positioned on each of the disk ansae, and the remaining four pointings were spaced evenly on either side of the ring.  All of these seven pointings were observed within a single 45 minute scheduling block (SB), which was executed four times on 2015 December 29-30 with 38 antennas in the array and an average precipitable water vapor (pwv) of $\sim0.75$~mm.  An additional three executions were carried out on 14 January 2016 with $44-46$ antennas in the array and pwv $\sim2.4$~mm.  Table~\ref{tab:obs} summarizes these observations including the dates, baseline lengths, weather conditions, and time on-source.  The two-week difference between observations produces a negligible pointing difference due to proper motion compared with the natural weight beam size, which we ignore.    

The correlator set-up for these observations was designed to optimize the continuum sensitivity, while also covering the $^{12}$CO J $=2-1$ transition at 230.538~GHz.  To achieve this, four basebands were centered at 213.98, 215.98, 229.59, and 231.48~ GHz, in two polarizations.  The baseband covering the $^{12}$CO spectral line included 3840 channels over a bandwidth of 1.875 GHz with a velocity resolution of 1.27 km/s.  The remaining three continuum basebands included only 128 channels with a total bandwidth of 2 GHz each.

The raw datasets were processed by ALMA staff using the \texttt{CASA} software package (version 4.5.2).  The absolute flux calibration was determined from observations of Pallas, J2357-5311, and J2258-275, with a systematic uncertainty of $<10\%$.  Observations of J2258-2758 were also used to determine the bandpass calibration and to account for time-dependent gain variations.  To reduce the size of the dataset, the visibilities were averaged into 30 second intervals.  We generated continuum images using the multi-frequency synthesis \texttt{CLEAN} algorithm in \texttt{CASA}, and correct for the telescope primary beam.  In Band 6, the primary beam of the ALMA 12-m antennas has FWHM$\sim26\arcsec$.  The imaging of the $^{12}$CO data is described in Matr\`{a} et al. (2017).

\section{Results and Analysis}
\label{sec:results}

\subsection{Continuum Emission}
\label{sec:cont}

\begin{figure}[ht]
\begin{minipage}[h]{0.49\textwidth}
  \begin{center}
       \includegraphics[scale=0.73]{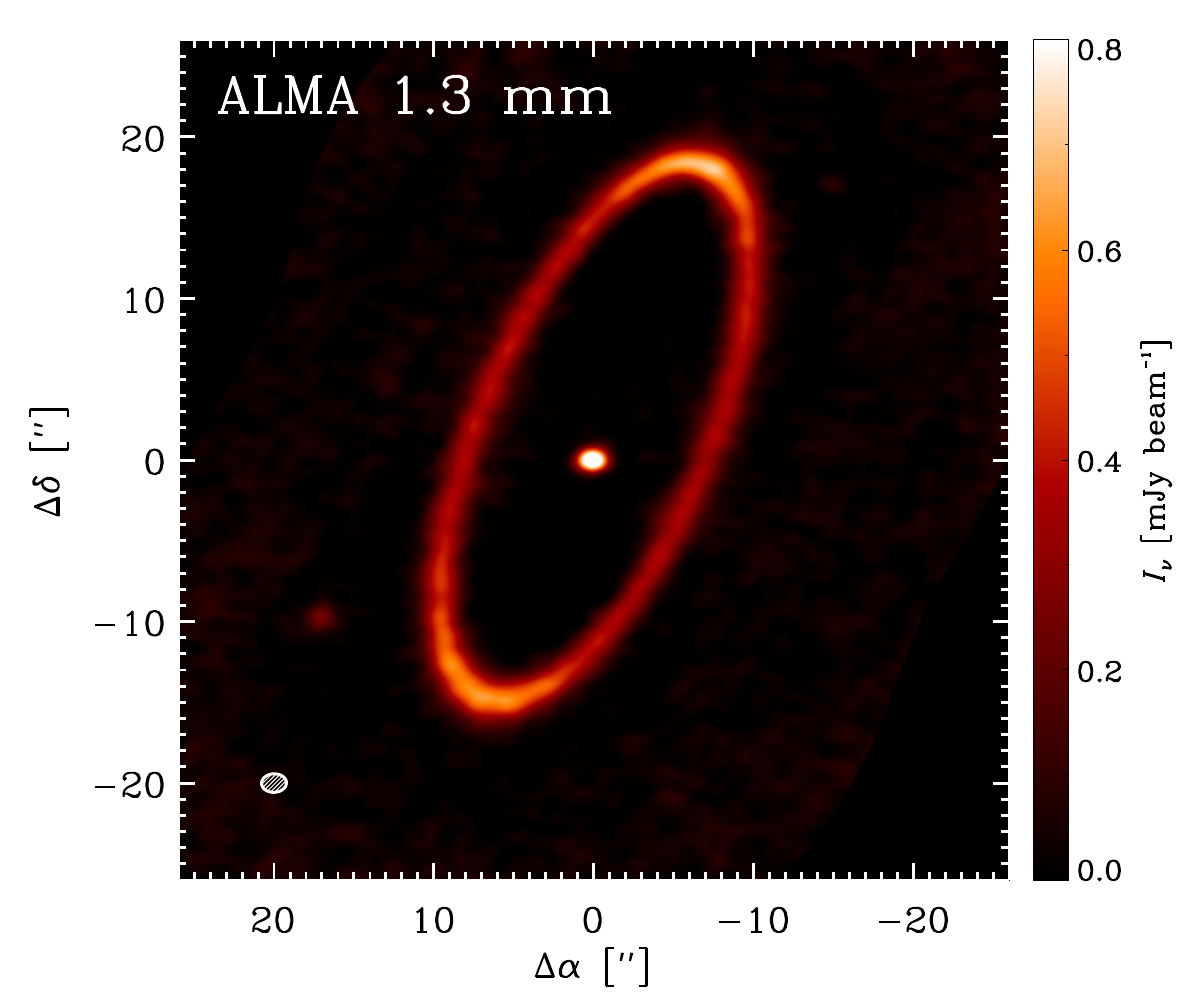}
  \end{center}
 \end{minipage}
\begin{minipage}[h]{0.49\textwidth}
  \begin{center}
       \includegraphics[scale=0.73]{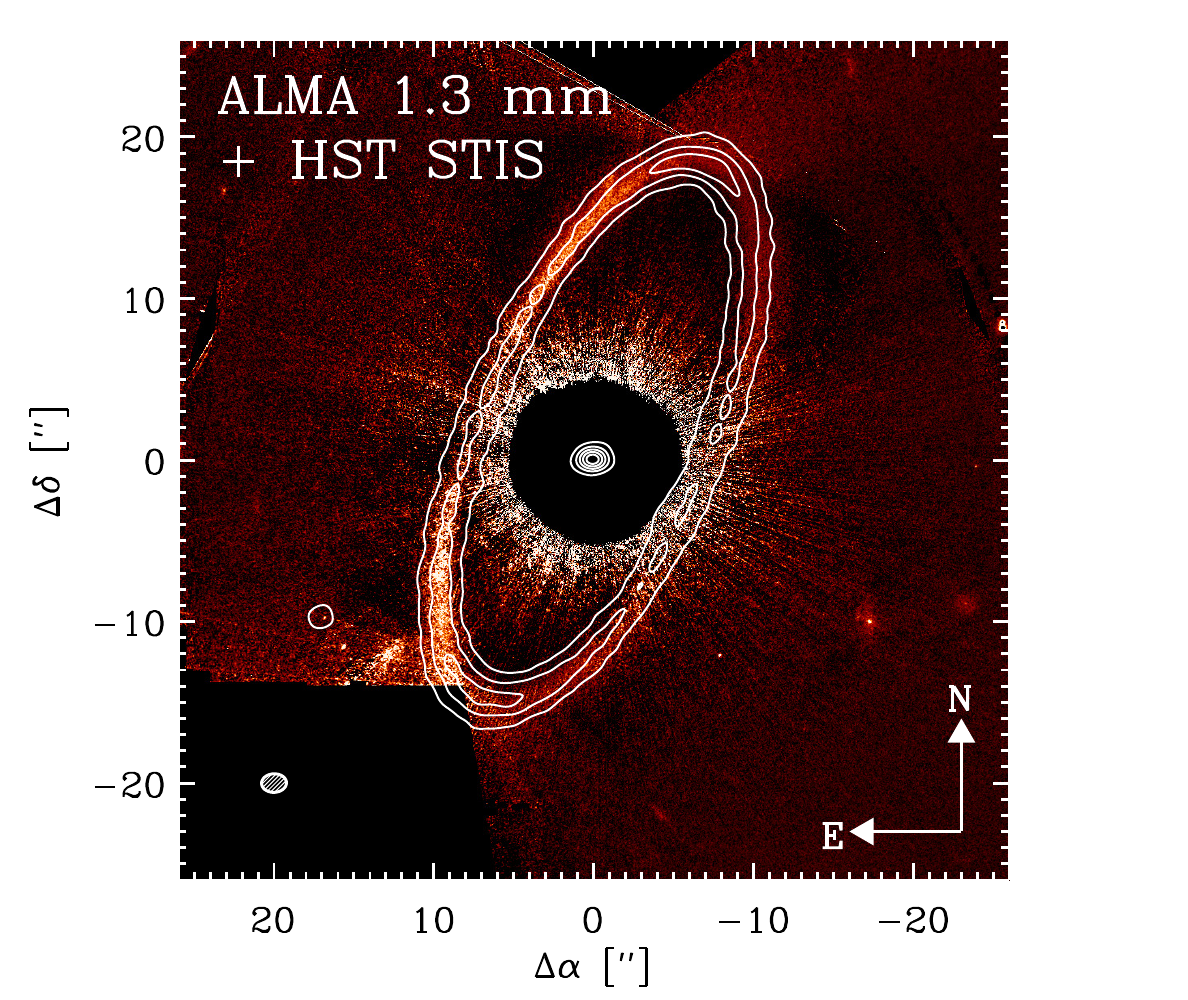}
  \end{center}
 \end{minipage}
 
\caption{\small  \emph{(left)} ALMA image of the 1.3~mm continuum emission from Fomalhaut.  The dashed white ellipse in the lower left corner shows the natural weight beam size of $1\farcs56\times1\farcs15$.  The rms noise is 14~$\mu$Jy/beam.
\emph{(right)} The ALMA continuum image overlaid as contours (white) on the HST STIS image from \cite{kal13}.  Contour levels are in steps of $[5,15,25,35,45,55]\times$ the rms noise.
}
\label{fig:cont}
\end{figure}

Figure~\ref{fig:cont} (left panel) shows the primary beam corrected ALMA 1.3~mm continuum image of Fomalhaut.  With natural weighting, the rms noise level is 14~$\mu$Jy/beam and the synthesized beam size is  $1\farcs56\times1\farcs15$ ($12\times9$~AU at 7.7~pc) with a position angle of $-87\degr$.  The right panel of Figure~\ref{fig:cont} shows the ALMA 1.3~mm image overlaid as contours on a \emph{Hubble
Space Telescope} (HST) STIS coronographic image of optical scattered light \citep{kal13}.  The millimeter continuum emission structure appears to match well with the narrow belt structure observed in the previous HST image.  Overall, the new ALMA image shows emission from three components: (1) a narrow, eccentric ring ($30\sigma$), (2) an unresolved central point source at the stellar position ($54 \sigma$), and (3) an unresolved point source on the eastern side of the disk ($10\sigma$).  Most strikingly, we note a significant flux difference between the apocenter (NW) and pericenter (SE) sides of the disk of $\sim65$~$\mu$Jy ($>5\sigma$), which we attribute to `apocenter glow,' a result of the disk's eccentricity \cite[][see Section~\ref{sec:apo} for further discussion]{pan16}.

We attribute the unresolved point source in the southeast quadrant to a background galaxy. The total flux density for this source is $0.150\pm 0.014$~mJy, determined by fitting a point source model to the visibilities using the \texttt{uvmodelfit} task in \texttt{CASA}.  Recent deep ALMA surveys have built up statistics on the number of faint background sources expected in a given field of view \citep{hat13,car15}. Given these (sub)millimeter source counts, the number of sources with flux density of $>0.15$~mJy expected within our field of view is $2.6^{+5.7}_{-1.9}$. The measured position of this point source is $\alpha=22^\text{h}57^\text{m}40.766$, $\delta=-29\degr37\arcmin32\farcs309$ (J2000).  This region has been imaged with HST/STIS in the optical (GO-13726; PI Kalas) where the nearest background source is $0\farcs68$ west and $0\farcs03$ north of the ALMA position.  Given that the ALMA beam radius is $\sim0\farcs78$ along Right Ascension, it is likely the ALMA source is the same background object as observed in optical data.

\subsection{Modeling Approach}
\label{sec:model}

Given the clear observed eccentricity in the Fomalhaut debris disk, we construct models that account for the orbital parameters of particles in the disk.  A particle orbiting within a circumstellar disk has both a proper and forced eccentricity, $e_p$ and $e_f$, respectively, as well as a proper and forced argument of periastron, $\omega_p$ and $\omega_f$.  We begin by populating the complex eccentricity plane defined by these four parameters following \cite{wya99}.  The forced eccentricity and argument of periastron, $e_f$ and $\omega_f$, are imposed on the particles by the massive perturber forcing the eccentricity in the disk, and are free parameters in our model.  The proper eccentricity is also left as a free parameter, $e_p$, and describes the additional scatter in the eccentricity of each particle's orbit; the $\omega_p$ associated with a given $e_p$ is assumed to be randomly distributed from $0$ to $2\pi$.  By assuming a semi-major axis, $a$, for each particle and random mean anomalies, we iterate to find the true anomaly, $f$, using the \texttt{newtonm} code from \texttt{ast2body} \citep{val07}.  Then, the radial orbital locations of each particle can be found simply using

\begin{equation}
r = \frac{a(1-e^2)}{1+e\text{cos}(f)}.
\end{equation}

\noindent To create our two-dimensional model, we complete this calculation for $10^4$ individual disk particles.  By creating a two-dimensional model, we assume that the disk structure has a negligible vertical component.  This assumption is motivated by the result from \cite{bol12} that the vertical scale height of the disk is described by an opening angle of $\sim1\degr$ from the mid-plane.  Adding a vertical component to the model would likely loosen the constraints we are able to place on the width of the belt (see Section~\ref{sec:width} for further discussion).  

To create an image, we bin the determined orbital locations into a two dimensional histogram with the bin size equal to the desired pixel scale and impose a radial temperature profile, $T\propto r^{-0.5}$.  The belt semi-major axis ($R_\text{belt}$) and range of semi-major axes ($\Delta a$), are both free parameters. In this eccentric disk model, the belt semi-major axis is the mean inner edge location, $R_\text{belt}=(R_\text{per}+R_\text{apo})/2$, where $R_\text{per}$ and $R_\text{apo}$ are the radial location of the disk inner edge at pericenter and apocenter, respectively.  The total flux density of the disk is normalized to $F_\text{belt} = \int I_\nu d \Omega$.  A point source with flux density, $F_\text{star}$, is added to account for the central stellar emission. In addition to fitting for both fluxes, we fit for the geometry of the disk (inclination, $i$, and position angle, $PA$), as well as offsets of the stellar position from the pointing center of the observations ($\Delta\alpha$ and $\Delta\delta$).

For a given model image, we compute synthetic model visibilities using \texttt{vis\textunderscore sample} \footnote{\texttt{vis\textunderscore sample} is publicly available at \url{https://github.com/AstroChem/vis_sample} or in the Anaconda Cloud at \url{https://anaconda.org/rloomis/vis_sample}}, a python implementation of the Miriad  \texttt{uvmodel} task.  Following our previous approach \cite[e.g.][]{mac13,mac16b}, we evaluate these model visibilities using a $\chi^2$ likelihood function that incorporates the statistical weights on each visibility measurement.  This iterative process makes use of the \texttt{emcee} Markov Chain Monte Charlo (MCMC) package \citep{for13}.  Given the affine-invariant nature of this ensemble sampler, we are able to explore the uncertainties and determine the one dimensional marginalized probability distribution for each independent model parameter.

\subsection{Results of Model Fits}
\label{sec:fits}

\begin{figure}[ht]
  \begin{center}
       \includegraphics[scale=1]{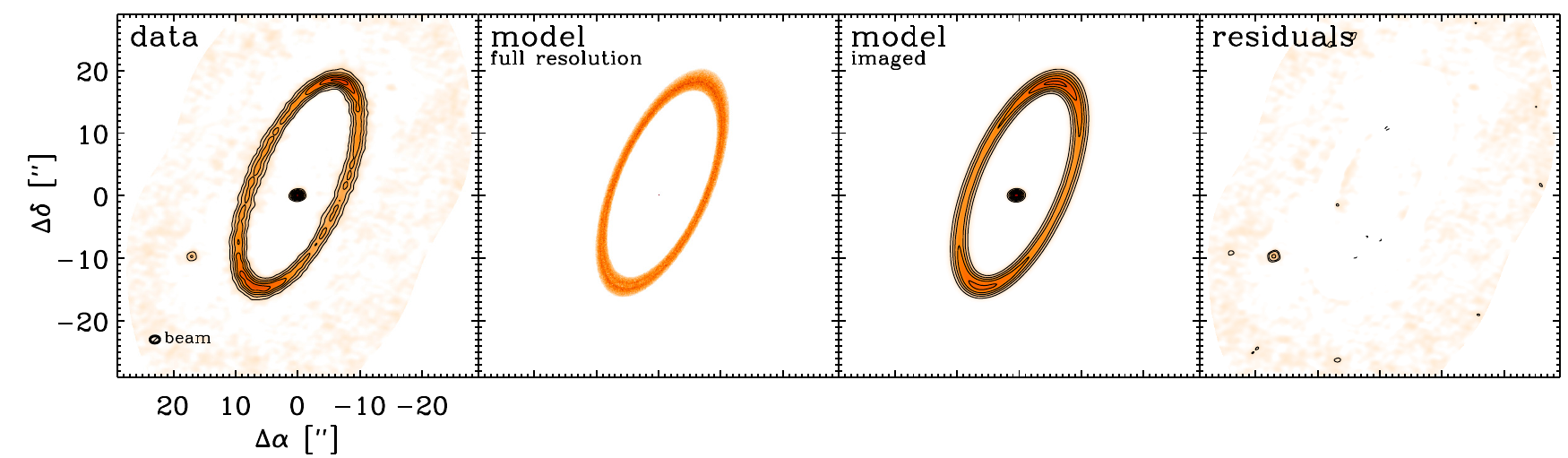}
  \end{center}
\caption{\small \emph{(left)} Image of the ALMA 1.3~mm continuum emission. \emph{(left, center)} The best-fit model at full resolution with pixel scale $\sim0.1\arcsec$ ($\sim0.8$~AU). \emph{(right, center)} The best-fit model imaged like the data with no noise. \emph{(right)} The residuals of the best-fit model with the same imaging parameters.  Contour levels in the first three panels are in steps of $5\times$ the rms noise of 14~$\mu$Jy/beam.  In the rightmost panel, additional contours of $\pm3\sigma$ are added to highlight any residual structure.  The labeled ellipse in the lower left corner indicates the synthesized beam size of $1\farcs56\times1\farcs15$, the same as in Figure~\ref{fig:cont}.
}
\label{fig:dmr}
\end{figure}

Table~\ref{tab:diskpar} presents the best-fit model (reduced $\chi^2 = 1.1$) parameters and their $1\sigma$ ($68\%$) uncertainties.  Figure~\ref{fig:dmr} shows the ALMA 1.3 mm data (left panel) along with the best-fit model displayed at full resolution and imaged like the ALMA data (center panels).  The rightmost panel shows the imaged residuals resulting from subtracting this best-fit model from the data, which are mostly noise.  The only significant peak corresponds to the background galaxy discussed in Section~\ref{sec:cont}. The full MCMC output is shown in the appendix.

The total belt flux density determined for the best-fit model is $24.7\pm0.1$~mJy (with an additional $10\%$ uncertainty from flux calibration), consistent with previous flux measurements at slightly shorter wavelengths.  \cite{bol12} determine a total flux density at 860~$\mu$m of $\sim85$~mJy, estimated from ALMA observations of the NW half of the ring.  \cite{hol98} and \cite{hol03} determine flux densities of $81\pm7.2$~mJy and $97\pm5$~mJy from SCUBA imaging at 450 and 850~$\mu$m, respectively.  Assuming a millimeter spectral index of $\sim2.7$ \citep{ric12}, the measurement from \cite{bol12} extrapolates to $\sim27$~mJy at 1.3~mm, consistent with our results within the uncertainties.  Using ALMA observations at 233~GHz ($\sim1.3$~mm), \cite{whi16} obtain a flux density of $30.8^{+3.4}_{-1.0}$~mJy by fitting directly to the visibilities and $26.3^{+4.5}_{-4.7}$~mJy  by fitting in the image plane, again consistent with our results within the mutual uncertainties. For optically thin dust emission, the total dust mass is given by $M_\text{dust} = F_\nu D^2/(\kappa_\nu B_\nu(T_\text{dust}))$, where $D=7.66$~pc is the distance, $B_\nu(T_\text{dust})$ is the Planck function at the dust temperature, $T_\text{dust}$, and $\kappa_\nu$ is the dust opacity.  We assume a dust opacity at 1.3~mm of $\kappa_\nu = 2.3$ cm$^{2}$ g$^{-1}$ \citep{beck90}, which may be a source of systematic uncertainty.  Given the best fit radial location of the disk, $136.3\pm0.9$~AU, the radiative equilibrium temperature is $\sim48$~K.  Thus, the total mass of the dust belt is $0.015\pm0.010$~$M_\oplus$.

We find good agreement with all previous determinations between the belt semi-major axis, eccentricity, inclination, and position angle for our best-fit model with previous results.  The best-fit belt semi-major axis from our modeling is $136.3\pm0.9$~AU.  At pericenter, the inner edge of the belt is located at a radial distance of $R_\text{per} = (1-e)R_\text{belt} = 119.9\pm0.8$~AU.  At apocenter, the inner edge of the belt is at $R_\text{apo} = (1+e)R_\text{belt} = 152.6\pm1.0$~AU.  HST imaging yields a value of $136.28\pm0.28$~AU \citep{kal05,kal13}, while \cite{acke12} obtain $137.5\pm0.9$~AU from \emph{Herschel} observations.  \cite{bol12} determine a semi-major axis of $135^{+1.0}_{-1.5}$ from ALMA imaging of the NW half of the disk, and \cite{whi16} determine a belt center location of $139^{+2}_{-3}$~AU from their model fits. These same observational studies yield inclination and position angles that range from $65\degr - 67\degr$ and $336\degr - 350\degr$, respectively.  We obtain robust constraints on both angles of $i = 65\fdg6\pm0\fdg3$ and $PA = 337\fdg9\pm0\fdg3$.  The best-fit eccentricity is $0.12\pm0.01$, consistent with both the \emph{Herschel} result of $0.125\pm0.006$ and with the HST and previous ALMA results of $0.11\pm0.01$.

\section{Discussion}
\label{sec:disc}

For the first time, we have resolved the complete Fomalhaut outer debris disk at 1.3~mm with ALMA.  This map of the dust continuum emission reveals a narrow, eccentric ring surrounding the primary star.  Apocenter glow, a result of increased surface density at apocenter in an eccentric ring, is evident as a significant brightness difference between the NW and SE sides of the disk.  Our modeling results place strong constraints on the disk position, width, geometry (inclination and position angle), eccentricity, and argument of periastron.  We now use these new results to discuss implications for the grain composition, azimuthal structure of the disk, the directly imaged object interior to the disk, Fomalhaut b, and the central star.

\subsection{Observational Evidence for Apocenter Glow}
\label{sec:apo}

Our new ALMA image is the first conclusive observational evidence for apocenter glow.  The Keplerian orbital velocity in an eccentric disk is slower at apocenter than at pericenter producing an overdensity of material at apocenter.  At mid-infrared wavelengths, the observed flux is strongly dependent on the grain temperature; grains at pericenter glow more brightly since they receive more flux from the star, masking the apocenter overdensity.  This effect is evident as `pericenter glow' \citep{wya99} in \emph{Herschel} images of the Fomalhaut disk at 70~$\mu$m, where the SE (pericenter) side of the disk appears brighter \citep{acke12}.  In contrast, previous imaging of the Fomalhaut debris disk at longer far-infrared to millimeter wavelengths suggests a slight excess ($<3\sigma$) of emission at the NW (apocenter) side of the disk, farthest from the star \citep{hol03,mar05,ric12}. To explain this phenomenon, \cite{pan16} construct a model of `apocenter glow' where the enhancement of the surface density of the disk at apocenter results in wavelength-dependent surface brightness variations.   At millimeter wavelengths, larger grains dominate the emission.  Since these grains radiate efficiently at the blackbody peak, the pericenter-apocenter temperature difference has less impact on the total flux.  As a result, the larger surface density at apocenter dominates and the apocenter appears brighter.  

\begin{figure}[ht]
  \begin{center}
       \includegraphics[scale=0.55]{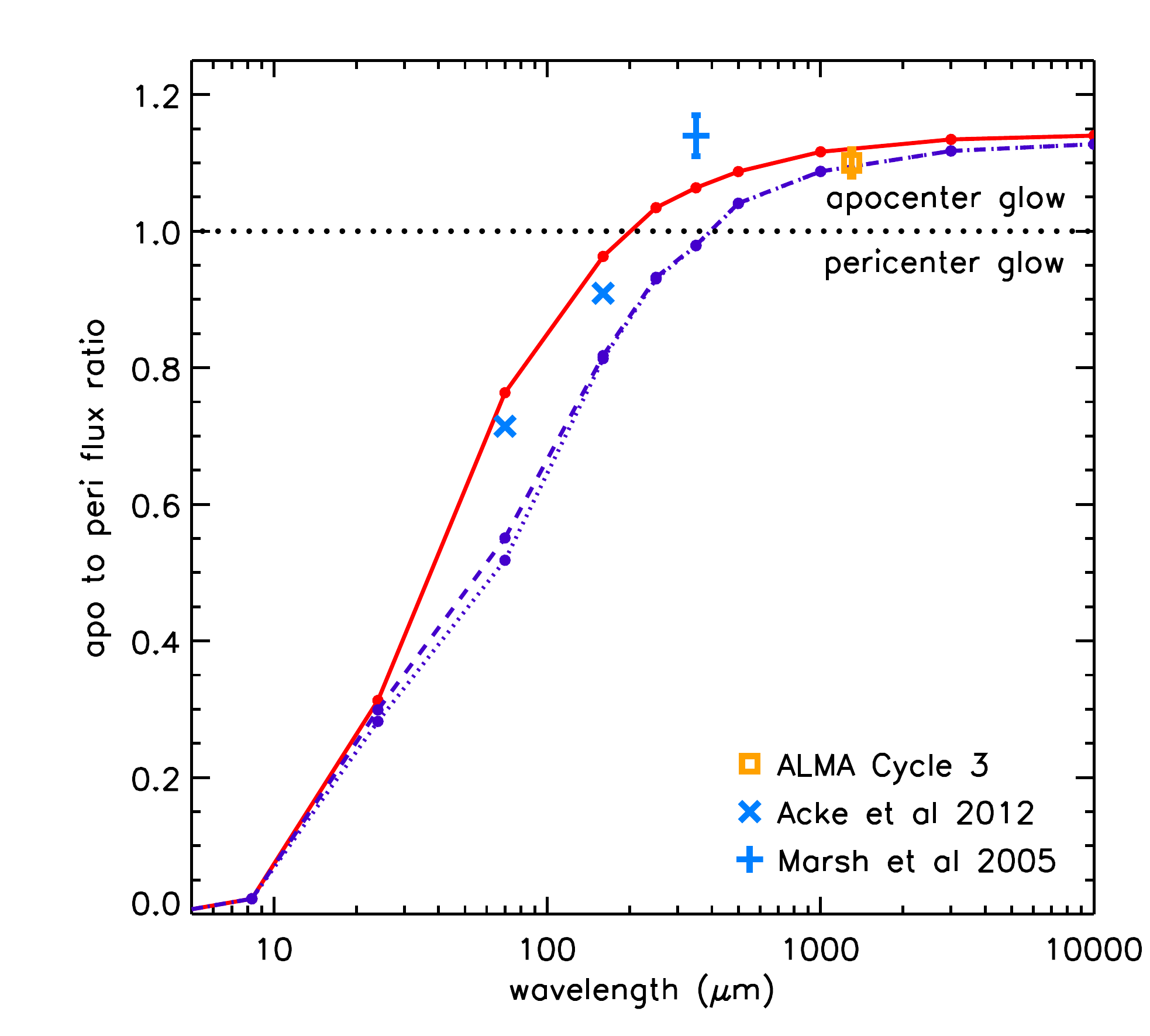}
  \end{center}
\caption{\small Apocenter to pericenter flux ratios (ratio of the radially integrated disk flux at apocenter to that at pericenter) as a function of wavelength. The yellow square indicates our new ALMA measurement. Blue points are measured flux ratios from {\em Herschel} observations at 70 and 160~$\mu$m \citep{acke12} and from CSO/SHARC II observations at 350~$\mu$m \citep{mar05}. Uncertainties on the {\em Herschel} points are smaller than the plot symbols. The curves outline the region obtained in our grid of Fomalhaut disk simulations. The red solid curve follows the maximum flux ratio values, which occur at $q=4$, $\beta=1$; the purple dashed/dotted curves follow the flux ratios occurring at $q=3$ and $\beta=2$ (dashed) or $\beta=3$ (dotted). The $q=3$, $\beta=3$ flux ratios are the minimum attained on our parameter grid: extending our upper bound on $\beta$ from 2 to 3 makes little difference in the overall range of model flux ratios. The observed results show broad agreement with our simulations. 
}
\label{fig:apo}
\end{figure}

Figure~\ref{fig:apo} shows the apocenter to pericenter flux ratio for Fomalhaut as a function of wavelength, including our new ALMA measurement at 1.3~mm of $1.10\pm0.02$. Plotted together with the observational results are curves showing the smallest (purple dotted line) and largest (red solid line) apocenter to pericenter flux ratios obtained with a grid of simulated Fomalhaut disks. A detailed description of the disk simulations is given by \cite{pan16}; here, we include a brief overview. We created disks with the forced eccentricity $e_f$, radial location $R_\mathrm{belt}$, and semimajor axis range $\Delta a$ given in Table~\ref{tab:diskpar} orbiting stars with effective temperature $T_*=8590$~K and radius $R_*=1.28\times 10^{11}$~cm \citep{mam12}. We populated the disks with particles of sizes $a$ following power-law size distributions $\text{d}n/\text{d}a\propto r^{-q}$ and grain absorptivities $Q\propto a^{-\beta}$. We drew each model disk's $q$ and $\beta$ values from a grid covering the ranges $3\leq q\leq 4$, $1\leq\beta\leq 3$. We then calculated the radially-integrated disk brightness as a function of longitude assuming passively heated, optically thin disks in thermal equilibrium. The ALMA flux ratio measurement falls well within the range obtained in our model grid.

As Figure~\ref{fig:apo} suggests, the observed apocenter to pericenter flux ratios can be diagnostic of disk grain properties including $\beta$, the grain absorptivity, and $q$, the size distribution power law index.  The long wavelength spectral index, $\alpha_\text{mm}$, of dust emission constrains the size distribution of dust grains in the disk.  Again assuming that the differential number of grains of size $a$ is a power law, $\text{d}n/\text{d}a\propto a^{-q}$, then $q = (\alpha_\text{mm} - \alpha_\text{Pl})/\beta_s + 3$ \citep{ric12,mac16a}.  Here, $\alpha_\text{Pl}=1.88\pm0.02$ \cite[see discussion in][]{mac16a}, and $\beta_s = 1.8\pm0.2$, the dust opacity spectral index in the small particle limit for interstellar grain materials \citep{dra06}. We note that different assumptions for the dust opacity can produce steeper grain size distributions \citep{gas12}.  \cite{ric12} measured the flux density of Fomalhaut at 6.66~mm with ATCA.  By pairing our new ALMA flux density with this previous measurement, we determine $\alpha_\text{mm} = 2.71\pm0.11$ and thus $q = 3.46\pm0.09$. This result is consistent with the determination of \cite{whi16} of $\alpha_\text{mm}=2.73\pm0.13$ and $q=3.50\pm0.14$.  Using the flux ratios measured at 70~$\mu$m, 160~$\mu$m, and 1.3~mm respectively, and with a slight extension in the parameter range for our models, our $1\sigma$ uncertainty range in $q$ implies $0.9<\beta<1.6$, $0.7<\beta<1.5$, and $0.7<\beta$.\footnote{Extending our parameter grid range up to $\beta=4$ did not increase the range of flux ratios attained in our models enough to fix an upper bound on the $\beta$ values using the 1.3~mm data point.} The overlap between these indicates an allowed range of $0.9<\beta<1.5$, consistent within $1\sigma$ with the $\beta\simeq (q-3)\beta_s$ quoted by \citet{dra06}.

\subsection{Structure of Fomalhaut's Outer Debris Disk}
\label{sec:struc}

\subsubsection{Constraints on Azimuthal Belt Structure}
\label{sec:azi}

\begin{figure}[ht]
  \begin{center}
       \includegraphics[scale=1]{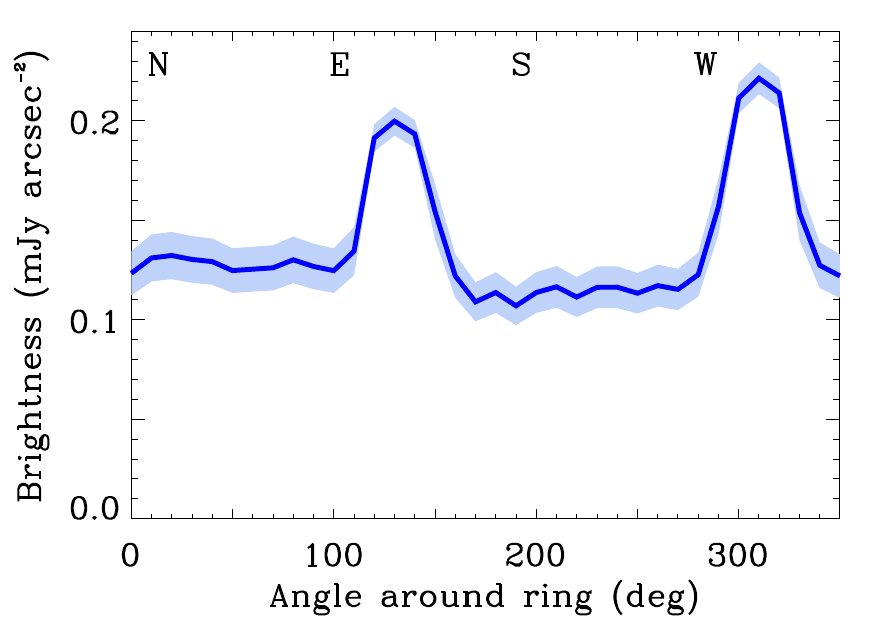}
  \end{center}
\caption{\small Azimuthal profile of the ALMA 1.3~mm continuum emission generated by calculating the mean brightness in $10\degr$ annular sections around the disk counterclockwise from North to East.  The disk ansae are clearly seen as two peaks, and apocenter appears brighter due to the detected apocenter glow.  The shaded region indicates the $\pm1\sigma$ confidence interval.
}
\label{fig:azi}
\end{figure}

The ALMA 1.3~mm mosaic map of the outer Fomalhaut debris disk was designed to cover the complete ring with equal sensitivity, allowing us to examine azimuthal structure along the belt.  After subtracting our best-fit model, any azimuthal structure should be clearly visible in the imaged residuals.  Figure~\ref{fig:dmr} shows the resulting residuals and no significant peaks are visible along the disk.  Figure~\ref{fig:azi} shows the azimuthal profile of the disk in the sky-plane.  The mean brightness is calculated in small annular sections of $10\degr$ around the ring starting in the North and moving counterclockwise to the East. Uncertainties are obtained by dividing the rms noise of the image by the square root of the number of beams in each annular sector.  The two disk ansae are visible as two peaks in the SE and NW, and apocenter glow is indicated by the significant brightness difference between these two peaks. No other significant peaks or fluctuations are present.  We note a slight brightness difference ($<3\sigma$) between the NE and SW sides of the disk (along the direction of the disk minor axis).  The median belt flux density measured between $170\degr-270\degr$ (SW side) is $0.11\pm0.01$~mJy~arcsec$^{-2}$ and $0.13\pm0.01$~mJy~arcsec$^{-2}$ between $0\degr-100\degr$ (NE side).  A similar dimming of the SW side of the disk is seen by \cite{bol12}, which they interpret as resulting from a loss of sensitivity at the edges of the ALMA primary beam.  However, it is likely that this slight asymmetry between the NE and SW sides of the disk presents further evidence for apocenter glow.  The expected overdensity of particles at apocenter forms an arc, which would cover much of the eastern side of the disk given the observed disk geometry \citep{pan16,pea14}.

Previous imaging surveys at optical to infrared wavelengths have indicated several azimuthal features, which our millimeter observations do not confirm.  \cite{kal13} demonstrate that the dust belt has a $\sim50\%$ deficit of optical scattered light in an azimuthal wedge at position angle $\sim331\degr$, just north of the current location of Fomalhaut b.  The sky-plane width of the gap is $2\arcsec$ ($\sim15$~AU), corresponding to a deprojected width of $\sim50$~AU.  One possibility is that the gap in scattered light represents a deficit of material, where grains collect on horseshoe orbits on either side of a planet embedded in the gap.  The brightness deficit could also result from self-shadowing in an optically thick, vertically thin belt.  Millimeter observations are minimally affected by optical depth effects and should reveal the true surface density of grains.  Since our ALMA observations do not detect the same $331\degr$ gap, it is likely that this feature results from a shadowing effect. However, we cannot rule out smaller structures $\lesssim10$~AU that would remain unresolved in our current map.

SCUBA imaging at 450~$\mu$m shows evidence for an arc of emission at position angle $\sim141\degr$ interior to the outer belt at $\sim100$~AU separation from the star \cite{hol03}.  \cite{bol12} note a broadening of the disk width on the northwestern side of the belt, to the right of the disk ansae.  We do not confirm either of these features in our ALMA map. \cite{whi16} also note that the disk appears azimuthally  smooth.

\subsubsection{Determining the Belt Width}
\label{sec:width}

\begin{figure}[ht]
  \begin{center}
       \includegraphics[scale=0.95]{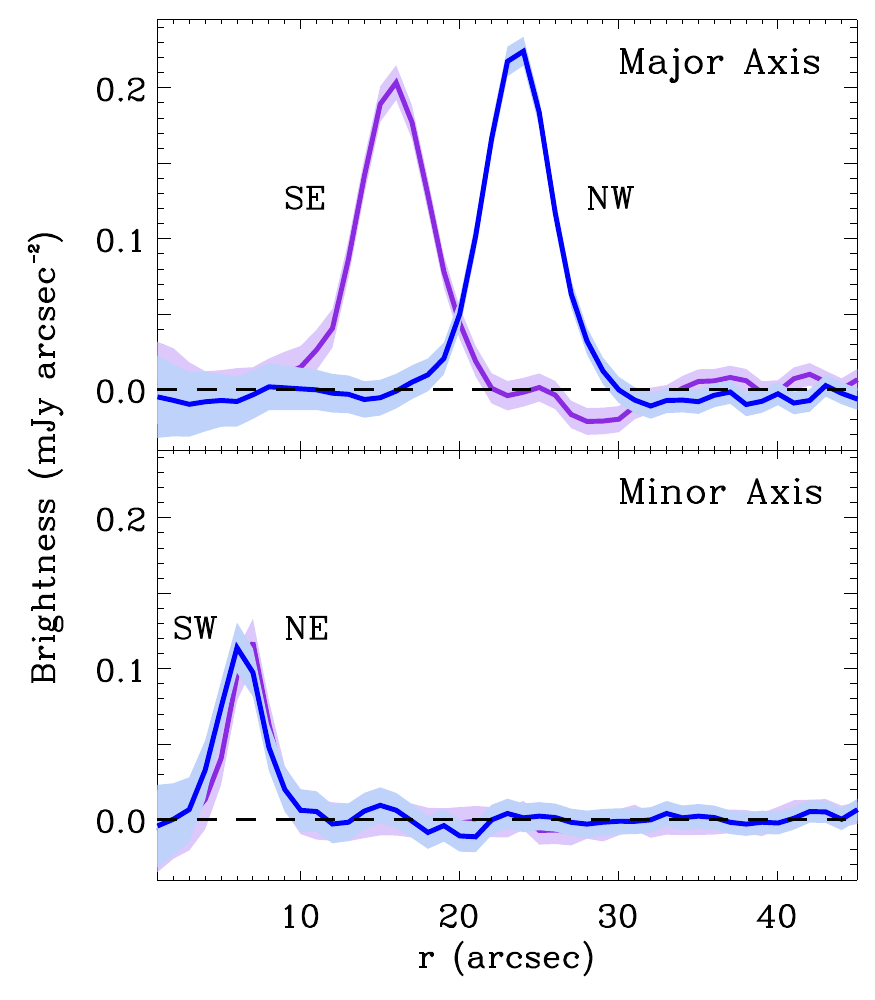}
  \end{center}
\caption{\small Surface brightness of the ALMA 1.3~mm continuum image in four cuts starting from the star: \emph{(top)} along the disk major axis to the NW (apocenter) and SE (pericenter) and \emph{(bottom)} along the disk minor axis to the SW and NE. The shaded regions indicate the $\pm1\sigma$ confidence interval.
}
\label{fig:slices}
\end{figure}

The FWHM width of our best-fit model is $13.5\pm1.8$~AU.  \cite{bol12} estimate a half-maximum width for the disk of $\sim11.4$~AU given a power-law belt model and $\sim16$~AU given a Gaussian model, consistent with our results.  \cite{whi16} determine a belt width of $13\pm3$~AU from their recent ALMA data. Figure~\ref{fig:slices} shows the surface brightness of our ALMA image in four cuts from the star along both the disk major (SE and NW sides) and minor (SW and NE) axes.  We do not see any fluctuation in width along the belt.  The flux difference between apocenter and pericenter is evident.  Also of note is the offset of the star from the disk centroid to the SW by $\sim0\farcs30$ ($\sim2.3$~AU) in RA and $\sim1\farcs4$ ($\sim10.7$~AU) in DEC.     

Given the best-fit parameters of our two-dimensional model, we can constrain the fractional width of the belt to be $\Delta R/R = 0.10\pm0.01$.  Adding a vertical component to the model likely adds to the uncertainty of this constraint.  The Fomalhaut debris disk is similarly narrow to the main classical Kuiper Belt in our own Solar System, which is radially confined between the 3:2 and 2:1 orbital resonances with Neptune implying a fractional width of $\sim0.18$ \citep{hahn05}.  In contrast, both the HD 107146 \citep{ric15} and $\eta$ Corvi \citep{mar17} debris disks appear much broader with fractional widths of $>0.3$.  \cite{bol12} propose that the narrow ring observed in Fomalhaut may also result from interactions with planets, namely two shepherding planets on the inner and outer edges of the belt.  If the structure of the belt is indeed due to truncation by interior and exterior planets, we would expect to see sharp edges.  However, given the resolution of our observations ($\sim10$~AU) compared with the width of the belt ($\sim14$~AU), we are unable to place any strong constraints on the sharpness of the disk edges.

\begin{figure}[ht]
  \begin{center}
       \includegraphics[scale=0.75]{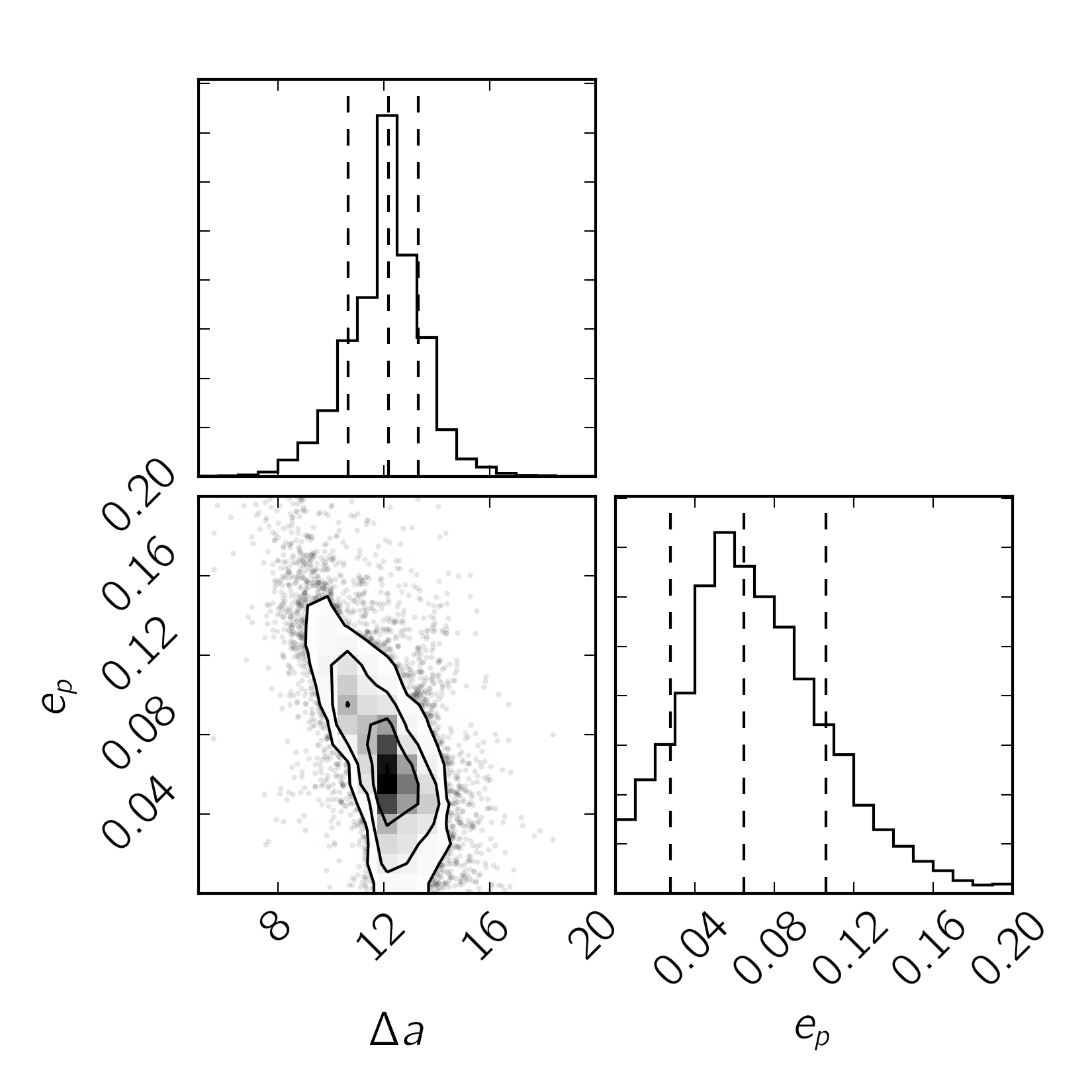}
  \end{center}
\caption{\small Results from $\sim10^4$ MCMC trials.  The diagonal plots show the 1D histogram for both $\Delta a$ and $e_p$ determined by marginalizing over the other parameter.  The dashed vertical lines indicate the best-fit value and $1\sigma$ uncertainty.  The off-diagonal plot shows the 2D projection of the posterior probability distribution for these two parameters.  Contours show the $1\sigma$, $2\sigma$, and $3\sigma$ regions.
}
\label{fig:ep}
\end{figure}

In our models, there are two parameters that contribute to the width of the belt: the range of semi-major axes assigned to the particles ($\Delta a$) and the proper or intrinsic eccentricity ($e_p$) of a particle's orbit.  As expected, these parameters are highly degenerate and we are unable to place strong constraints on either of these parameters independently given the moderate resolution of our observations.  The best-fit values for both parameters are $\Delta a = 12.2\pm1.6$~AU and $e_p = 0.06\pm0.04$.  Figure~\ref{fig:ep} shows the MCMC output for $\Delta a$ and $e_p$; the degeneracy between the two parameters is clearly seen by the slope in the contours.  Altering the proper eccentricity of the particles predicts azimuthal variations in the width of the belt.  For a low proper eccentricity ($e_p\sim0.01$), the particle orbits are apsidally aligned and the belt appears narrower at pericenter than at apocenter.  For a high proper eccentricity ($e_p\sim0.1$), the width of the belt is closer to uniform around the entire circumference of the ring.  Future ALMA observations of the disk apocenter and pericenter locations, but with higher resolution, could distinguish between these two cases, and place the first robust constraints on the proper eccentricity of the Fomalhaut debris disk.  Whereas \cite{whi16} have higher angular resolution in their recent ALMA observations (synthesized beam of $0\farcs329\times0\farcs234$), the two disk ansae are outside of the primary beam of their single ALMA pointing.

\subsubsection{Geometry of the Disk: The Argument of Periastron}
\label{sec:peri}

There has been much debate in the literature over the argument of periastron, $\omega_f$, of the Fomalhaut ring.  \cite{acke12} find $\omega_f=1\degr\pm6\degr$ based upon the location of the observed pericenter glow along the disk major axis.  However, the resolution of the \emph{Herschel} 70~$\mu$m image is not high enough to detect an offset in the stellar position off the disk major axis by a few AU.  \cite{kal13} determine a value of $29\fdg6\pm1\fdg3$ by fitting only for the offset of the expected stellar position from the disk centroid. \cite{bol12} are unable to constrain the argument of periastron, since they only image half of the belt with ALMA.

Our new ALMA data provides the first resolved image of emission from both the complete outer disk and the central star with high enough angular resolution to determine an offset of the star from the disk centroid.  As discussed in Section~\ref{sec:width}, the star is noticeably offset to the SW from the disk centroid.  This observation is consistent with the result of \cite{kal13}.  Adopting our modeling approach, we can fit independently for all three angles describing the disk geometry: the inclination ($i$), the position angle ($PA$), and the argument of periastron ($\omega_f$).  The best-fit argument of periastron from our models is $\omega_f = 22\fdg5\pm4\fdg3$.  This value is comparable to the result from \cite{kal13}, and matches both the stellar position relative to the disk centroid and the position along the belt of both the observed pericenter \citep{acke12} and apocenter glow.  There is still a large uncertainty in this best-fit value due to the difficulty disentangling the argument of periastron from the significant inclination of the disk ($i=65\fdg6\pm0\fdg3$).

\subsection{Implications for Fomalhaut b}
\label{sec:fomb}

Fomalhaut b was first discovered through HST direct imaging \citep{kal08} at a location consistent with theoretical predictions for a massive planet orbiting interior to the eccentric debris disk \citep{qui06,chi09}.  However, follow-up observations at later epochs revealed that Fomalhaut b is instead on a highly eccentric, possibly ring-crossing orbit \citep{kal13,beu14}.  Furthermore, this object appears brighter at optical wavelengths than in the infrared, contrary to predictions from models of planetary atmospheres.  \cite{ken11} discuss the possibility of a collisional swarm of irregular satellites surrounding a $\sim10$~$M_\oplus$ planet.  Alternatively, Fomalhaut b may instead be a dust cloud generated through collisions between larger planetesimals \citep{cur12,gal13,kal13,ken14,tam14,law15}.  To date, the true nature of Fomalhaut b remains uncertain.

If Fomalhaut b is indeed a dust cloud, our ALMA observations provide useful constraints on its possible dust mass. We can place a robust $3\sigma$ upper limit on the flux density of $0.042$~mJy, assuming a point source.  Following the approach for optically thin emission described in Section~\ref{sec:fits}, we can determine an upper limit on the potential dust mass.  The current separation of Fomalhaut b is $\sim125$~AU.  In radiative equilibrium, this location implies a dust temperature of $\sim51$~K.  The resulting upper limit on the dust mass is $<0.0019$~$M_\text{Moon}$ ($<1.40\times10^{23}$ g), which is consistent with estimates of the $10^{18} - 10^{21}$ g in total sub-micron dust mass needed to account for the scattered light \citep{kal08}.  We can also consider optically thick dust emission and instead derive an upper limit on the size of the dust clump: $R_\text{dust} = \sqrt{F_\nu D/(\pi B_\nu(T_\text{dust}))}$.  Given the upper limit of $F_\nu < 0.042$~mJy, $R_\text{dust}$ must be $<0.021$~AU for an optically thick clump.

\subsection{Stellar Emission at Millimeter Wavelengths}
\label{sec:star}

\begin{figure}[ht]
  \begin{center}
       \includegraphics[scale=0.95]{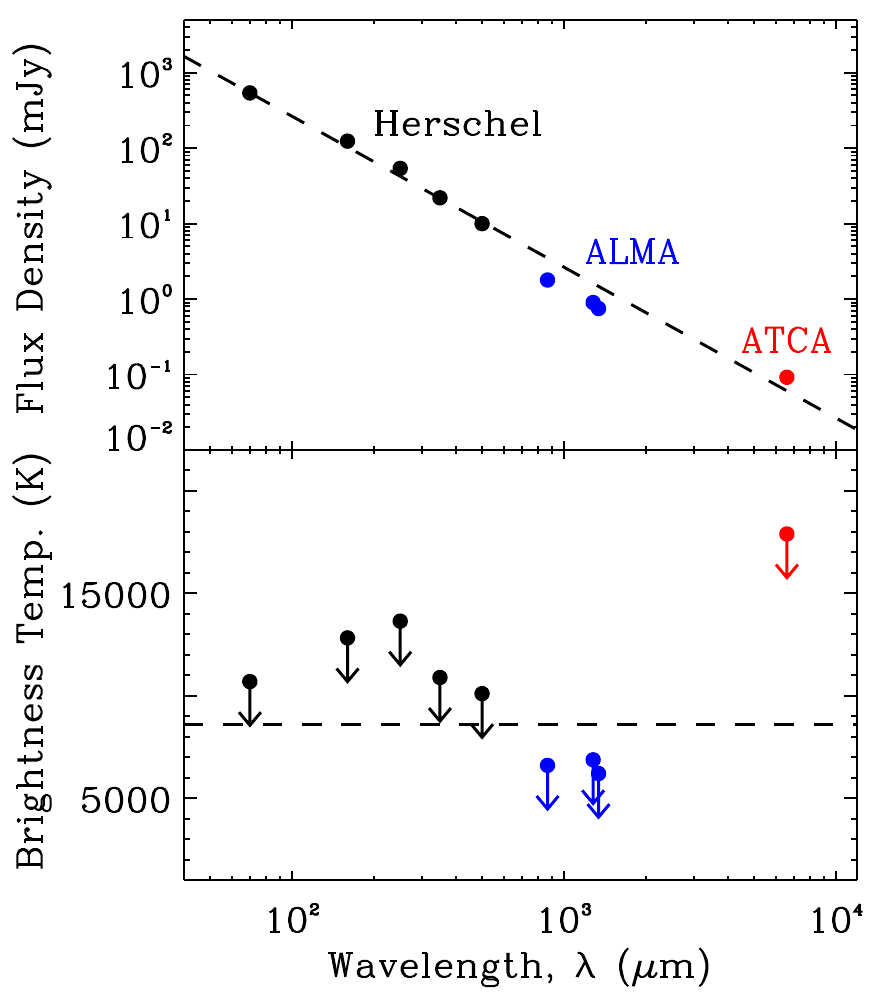}
  \end{center}
\caption{\small \emph{(top)} Flux density spectrum of Fomalhaut from \emph{Herschel} \cite[black points,][]{acke12}, ALMA \cite[blue points, this work;][]{su16,whi16}, and ATCA \cite[red point,][]{ric12}.  The dashed line indicates the expected spectral index for an optically thick photosphere with effective temperature $8590\pm73$~K.  The uncertainty on the flux measurements lies within the size of the points.  \emph{(bottom)}  Upper limits on the brightness temperature spectrum of Fomalhaut calculated by assuming a photospheric radius for the star.  Again, the dashed line indicates the expectation for a classical photosphere.  
}
\label{fig:star}
\end{figure}

The best fit flux density for the central star is $0.75\pm0.02$~mJy (with an additional $10\%$ uncertainty for flux calibration).  CHARA measurements of the stellar bolometric flux robustly determine the effective temperature to be $8459\pm44$~K \citep{boy13}.  Given this effective temperature, a PHOENIX stellar atmosphere model \citep{hus13} predicts a flux density of $\sim1.3$~mJy at 1.3~mm (with $5\%$ uncertainty), in excess of our flux measurement.  At long wavelengths, however, this stellar model is essentially a Rayleigh-Jeans extrapolation.  \cite{bol12} measure a stellar flux of $\sim4.4$~mJy at 850~$\mu$m with ALMA in Cycle 0, which extrapolates to $\sim1.8$~mJy at 1.3~mm, consistent with atmospheric model predictions, but not consistent with our ALMA flux. It is important to note, however, that this measurement is strongly influenced by the primary beam correction applied to the data, since the star is located at the edge of the single pointing.  ALMA Cycle 1 observations at 870~$\mu$m by \cite{su16} detect a central point source as well with a lower flux density of $1.789\pm0.037$~mJy.  Extrapolating to 1.3~mm, this measurement yields an expected flux density of $0.80\pm0.02$, more comparable to our result.  \cite{whi16} also determine a low stellar flux density of $0.90\pm0.15$~mJy from recent ALMA observations at a 1.3~mm (233~GHz, a slightly higher frequency than our observations). Figure~\ref{fig:star} shows the flux density spectrum (top) and brightness temperature (bottom) of Fomalhaut from \emph{Herschel} \citep{acke12}, ALMA \cite[this work;][]{su16,whi16}, and ATCA \citep{ric12}.  To calculate the brightness temperature, we follow \cite{lis16} and adopt a photospheric radius of $1.842\pm0.019$~$R_\odot$ \citep{mam12}. The stellar flux density at infrared wavelengths from \emph{Herschel} is inferred, since the measured flux includes contributions from both the star and the inner belt which are unresolved in these observations. Given the possible contribution from an inner warm belt at all wavelengths, we quote only upper limits on the brightness temperature. At the \emph{Herschel} wavelengths, the brightness temperature is mostly consistent with the effective temperature. However, at millimeter wavelengths, the brightness temperature dips to $<6600$~K and $<6200$~K at 870~$\mu$m and 1.3~mm, respectively.  The ATCA flux measurement at 6.66~mm suggests a brightness temperature of $<17900$~K.

It is clear that the brightness temperature of Fomalhaut is significantly lower than the measured photospheric effective temperature at millimeter wavelengths before increasing again at longer, centimeter wavelengths.  Similar behavior is seen for a number of K and M giants by \cite{har13}.  With the advent of ALMA, there are a growing number of stars with robust millimeter flux measurements.  Excess emission at long wavelengths has been observed for several other Sun-like stars, including $\alpha$ Cen A/B, $\epsilon$ Eridani, and $\tau$ Ceti \citep{lis15,lis16,mac15b,mac16b}, which is consistent with emission from a hot stellar chromosphere.  \cite{lis16} observe a temperature minimum, like we observe for Fomalhaut, for both $\alpha$ Cen A and B at shorter, sub-millimeter wavelengths with ALMA, which they attribute to a change in the sign of the temperature gradient above the stellar photosphere, as is seen in our own Sun.  At  1.3  mm  wavelength,  the  flux  densities  of  these  stars  have  recovered  and  their brightness temperatures are similar to their effective temperature.  For Fomalhaut, it seems likely that the flux density measured by ATCA at 6.66~mm is dominated by chromospheric emission.  However, the long wavelength spectrum of A-type stars, like Fomalhaut, is further complicated by ionized stellar winds, which flatten the spectral slope at radio wavelengths \citep{auf02}.  The ability to measure these behaviors with ALMA will enable advances in our understanding of stellar radiative transfer and chromospheres, and of stellar winds. 

For the Fomalhaut system, understanding the stellar flux contribution at long wavelengths is especially critical.  \emph{Spitzer} and \emph{Herschel} observations reveal excess emission at infrared wavelengths \citep{sta04,acke12}, which is attributed to a warm inner dust belt similar to the asteroid belt in our Solar System \citep{su13}. However, no inner belt has been detected or resolved with ALMA \citep{su16}.  Robustly determining the spectral energy distribution of the star at long wavelengths will help to determine the nature of such an inner asteroid belt.

\section{Conclusions}
\label{sec:conc}

We present new ALMA observations at 1.3~mm of the continuum dust emission from the Fomalhaut system.  These observations provide the first millimeter map covering the complete outer debris disk with uniform sensitivity.  We adopt a MCMC modeling approach that generates models of an eccentric ring by calculating the orbital parameters of a collection of particles.  The main results from this analysis are as follows.

\begin{enumerate}

\item The Fomalhaut outer debris disk is radially confined with a model best-fit inner edge of $136.3\pm0.9$~AU and width of $13.5\pm1.8$~AU, implying a fractional width of $0.10\pm0.01$.  Given the measured total flux density and assuming optically thin dust emission, the total dust mass of the disk is $0.015\pm0.010$~$M_\oplus$, consistent with previous measurements. Given the resolution of our observations, we do not place strong constraints on the sharpness of the belt edges.

\item Our ALMA image is the first conclusive observation of apocenter glow, a brightness asymmetry due to a surface density enhancement at apocenter \citep{pan16}.  We determine a best-fit eccentricity for the ring of $0.12\pm0.01$.  Given the apocenter to pericenter flux ratio from our ALMA measurement and previous (sub)millimeter observations, and assuming a size distribution power law index of $q = 3.46\pm0.09$, we constrain the dust absorptivity power law index $\beta$ to be $0.9<\beta<1.5$.

\item By adopting a modeling approach that accounts for the orbital parameters of disk particles, we are able to robustly constrain the geometry of the disk.  The best fit values for the inclination and position angle are $65\fdg6\pm0\fdg3$ and $337\fdg9\pm0\fdg3$, respectively. By resolving both the stellar position relative to the disk centroid and both the pericenter and apocenter sides of the disk, we are able to determine the argument of periastron to be $\omega = 22\fdg5\pm4\fdg3$, consistent with the results from HST images \citep{kal13}.

\item After subtracting our best-fit belt model from the data, the resulting residuals do not show any evidence for significant azimuthal structure.  The only significant peak visible to the east of the disk is attributable to a background galaxy.  We do not confirm any of the azimuthal features, including the gap at $331\degr$ position angle, that have been seen in previous imaging studies with HST, SCUBA, and ALMA.  However, we cannot rule out smaller structure $\lesssim10$~AU, which would be unresolved with the current resolution of our image.

\item The flux density at 1.3~mm of the central star, $F_\text{star} = 0.75\pm0.02$~mJy, is significantly lower than predicted by current photospheric models.  Indeed, the implied brightness temperature of Fomalhaut falls below the stellar effective temperature at millimeter wavelengths before increasing significantly at longer, centimeter wavelengths.  Similar spectra have been observed for the Sun-like stars $\alpha$ Cen A and B \citep{lis16}.  For Fomalhaut, it is especially critical to determine the long wavelength stellar spectrum in order to better constrain the contribution from the inner dust belt.

\end{enumerate}

The proximity (7.66~pc) and young age ($\sim440$~Myr) of the Fomalhaut system make it a unique target to explore the early stages of planetary system formation and reorganization.  Future ALMA observations with higher angular resolution will allow for further exploration of the outer disks's azimuthal structure, as well as enabling studies of structural variability over time.

\vspace{1cm}
M.A.M. acknowledges support from a National Science Foundation Graduate Research Fellowship (DGE1144152) and from NRAO Student Observing Support.  L.M. acknowledges support by STFC through a graduate studentship and, together with M.C.W. and A.S., by the European Union through ERC grant number 279973.  P.K. and J.R.G. thank support from NASA NNX15AC89G and NNX15AD95G/NEXSS, NSF AST-1518332 and HST-GO-13726.  This work benefited from NASA's Nexus for Exoplanet System Science (NExSS) research coordination network sponsored by NASA's Science Mission Directorate.  G.M.K. is supported by the Royal Society as a Royal Society University Research Fellow. M.P. acknowledges support from NASA grants NNX15AK23G and NNX15AM35G.  A.M.H. is supported by NSF grant AST-1412647. The work of O.P. is supported by the Royal Society Dorothy Hodgkin Fellowship.  A.S. is partially supported by funding from the Center for Exoplanets and Habitable Worlds. The Center for Exoplanets and Habitable Worlds is supported by the Pennsylvania State University, the Eberly College of Science, and the Pennsylvania Space Grant Consortium.  This paper makes use of the following ALMA data: ADS/JAO.ALMA \#2015.1.00966.S. ALMA is a partnership of ESO (representing its member states), NSF (USA) and NINS (Japan), together with NRC (Canada) and NSC and ASIAA (Taiwan) and KASI (Republic of Korea), in cooperation with the Republic of Chile. The Joint ALMA Observatory is operated by ESO, AUI/NRAO and NAOJ. The National Radio Astronomy Observatory is a facility of the National Science Foundation operated under cooperative agreement by Associated Universities, Inc.  Figures~\ref{fig:ep} and \ref{fig:appendix} were generated using the \texttt{corner.py} code \citep{for16}.

\bibliography{References}

\pagebreak

\begin{deluxetable}{c c c c c}
\tablecolumns{5}
\tabcolsep0.06in\footnotesize
\tabletypesize{\small}
\tablewidth{0pt}
\tablecaption{ALMA Observations of Fomalhaut \label{tab:obs}}
\tablehead{
\colhead{Observation} & 
\colhead{\# of } & 
\colhead{Projected} & 
\colhead{PWV} &
\colhead{Time on} \\
\colhead{Date} & 
\colhead{Antennas} & 
\colhead{Baselines (m)} &
\colhead{(mm)} &
\colhead{Target (min)}
}
\startdata
2015 Dec 29 & 38 & $15.1-310.2$ & 0.76 & 41.9 \\
 & 38 & $15.1-310.2$ & 0.65 & 41.9 \\
 & 38 & $15.1-310.2$ & 0.83 & 30.7 \\
2015 Dec 30 & 38 & $15.1-310.2$ & 1.1 & 26.6\\
2016 Jan 14 & 46 & $15.1-331.0$ & 2.3 & 41.9  \\
 & 46 & $15.1-331.0$ & 2.4 & 41.9 \\
 & 44 & $15.1-312.7$ & 2.7 & 41.9
\enddata
\end{deluxetable}

\begin{deluxetable}{l l c}
\tablecolumns{3}
\tabcolsep0.1in\footnotesize
\tabletypesize{\small}
\tablewidth{0pt}
\tablecaption{Best-fit Model Parameters \label{tab:diskpar}}
\tablehead{
\colhead{Parameter} & 
\colhead{Description} & 
\colhead{Best-fit Value}
}
\startdata
$F_\text{belt}$ & Total flux density [mJy] & $24.7\pm0.1$ \\
$F_\text{star}$ & Total stellar flux [mJy] & $0.75\pm0.02$ \\
$R_\text{belt}$ & Belt inner edge [AU] & $136.3\pm0.9$ \\
$\Delta a$ & Range of semi-major axes [AU] & $12.2\pm1.6$\\
$\Delta R$ & Belt FWHM [AU] & $13.5\pm1.8$ \\
$i$ & Disk inclination [$\degr$] & $65.6\pm0.3$ \\
$PA$ & Disk position angle [$\degr$] & $337.9\pm0.3$ \\
$e_f$ & Forced eccentricity & $0.12\pm0.01$ \\
$e_p$ & Proper eccentricity &  $0.06\pm0.04$\\
$\omega_f$ & Forced argument of periastron [$\degr$] & $22.5\pm4.3$ \\
$\Delta\alpha$ & RA offset [$\arcsec$] & $0.08\pm0.01$ \\
$\Delta\delta$ & DEC offset [$\arcsec$] & $0.06\pm0.01$ 
\enddata
\end{deluxetable}

\pagebreak

\appendix

\begin{figure}[ht]
  \begin{center}
       \includegraphics[scale=0.35]{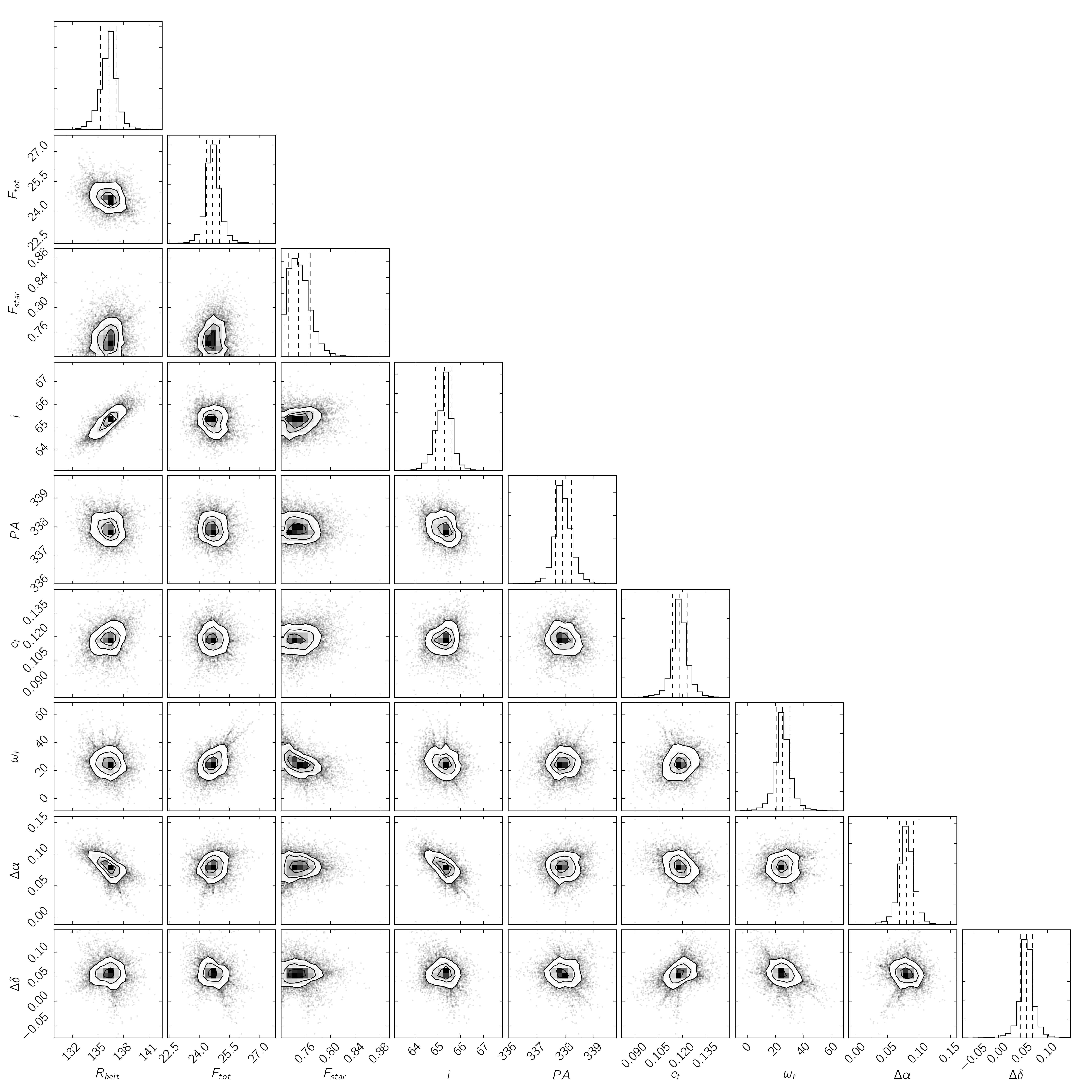}
  \end{center}
\caption{\small  
The 1D (diagonal panels) and 2D (off-diagonal panels) projections of the posterior probability distributions for the best-fit eccentric model parameters. For each parameter, the 1D histogram is determined by marginalizing over all other model parameters. The dashed vertical lines indicate the best-fit values and $1\sigma$ uncertainties (listed in Table~\ref{tab:diskpar}). The 2D joint probability distributions show the $1\sigma$, $2\sigma$, and $3\sigma$ regions for all parameter pairs.
}
\label{fig:appendix}
\end{figure} 

\end{document}